\documentclass[12pt,preprint]{aastex}

\newcommand{\K}{\,{\rm K}}

\newcommand{\simle}{\mbox{$\stackrel{<}{_{\sim}}$}}

\newcommand\arcdegg{\hbox{$^\circ$}}

\newcommand\etal{ {\em et~al.\/}\thinspace}

\newcommand\msun{\hbox{\,M$_\odot$}}
\newcommand\rsun{\hbox{\,R$_\odot$}}
\newcommand\lsun{\hbox{\,L$_\odot$}}
\newcommand\kms{ km~s$^{-1}$}

\newcommand\vis{${\mathcal{V}^2}$}
\newcommand\viss{${\mathcal{V}^2}$ }
\newcommand{\tpol}{T$_{\rm pol}$}
\newcommand{\rpol}{R$_{\rm pol}$}
\newcommand{\teq}{T$_{\rm eq}$}
\newcommand{\req}{R$_{\rm eq}$}

\newcommand\lbol{$\emph{L}_{\mathrm{bol}}$}

\newcommand\vsini{\rm{$v\sin i$} \ }
\newcommand\vsinii{\rm{$v\sin i$}}
\newcommand\wcrit{\rm{$\omega_{\rm crit}$}}
\newcommand\wratio{\rm{$\omega$ / \wcrit}}

\newcommand\teff{$\emph{T}_{\mathrm{eff}}$\ }
\newcommand\lapp{$\emph{L}_{\mathrm{app}}$\ }
\newcommand\tapp{$\emph{T}_{\mathrm{app}}^{\mathrm{eff}}$\ }

\shorttitle{Resolving Vega}
\shortauthors{Monnier et al.}

\begin{document}


\title{Resolving Vega and the inclination controversy with CHARA/MIRC}


\author{J.~D.~Monnier\altaffilmark{1}, Xiao Che\altaffilmark{1}, Ming Zhao\altaffilmark{2}, S. Ekstr\"om\altaffilmark{3}, 
V. Maestro\altaffilmark{4}, J. Aufdenberg\altaffilmark{5}, F. Baron\altaffilmark{1}, C. Georgy\altaffilmark{6}, S. Kraus\altaffilmark{1}, H. McAlister\altaffilmark{7}, E. Pedretti\altaffilmark{8}, S. Ridgway\altaffilmark{9}, 
J. Sturmann\altaffilmark{7}, L. Sturmann\altaffilmark{7}, T. ten Brummelaar\altaffilmark{7}, N. Thureau\altaffilmark{10},   N. Turner\altaffilmark{7}, P. G. Tuthill\altaffilmark{4}
}

\altaffiltext{1}{monnier@umich.edu: University of Michigan (Astronomy), 500 Church St, Ann Arbor, MI 48109}
\altaffiltext{2}{The Pennsylvania State University, University Park, PA 16802, USA}
\altaffiltext{3}{Geneva Observatory, University of Geneva, Maillettes 51, 1290, Sauverny, Switzerland}
\altaffiltext{4}{Sydney Institute for Astronomy (SIfA), School of Physics, University of Sydney, NSW 2006, Australia}
\altaffiltext{5}{Embry-Riddle Aeronautical Univ, Daytona Beach, FL 32114 USA }
\altaffiltext{6}{Centre de Recherche Astrophysique, Ecole Normale Sup\'erieure de Lyon, F-69384 Lyon cedex 07, France}
\altaffiltext{7}{The CHARA Array of Georgia State University, Mt. Wilson, CA, 91023, USA}
\altaffiltext{8}{The Scottish Association for Marine Science, Dunstaffnage, Oban, Argyll PA37 1QA UK}
\altaffiltext{9}{National Optical Astronomy Observatory, 950 N. Cherry Ave, Tucson AZ 85719. USA}
\altaffiltext{10}{Department of Physics and Astronomy, University of St. Andrews, Scotland UK }

\email{JDM: monnier@umich.edu}


\begin{abstract}
Optical and infrared interferometers definitively established that the photometric standard Vega ($=\alpha$~Lyrae)  is a rapidly rotating star viewed nearly pole-on.  Recent independent spectroscopic analyses could not reconcile the inferred inclination angle with the observed line profiles, preferring a larger inclination.  In order to resolve this controversy, we observed Vega using the six-beam Michigan Infrared Combiner (MIRC6) on the Center for High Angular Resolution Astronomy (CHARA) Array.  With our greater angular resolution and dense (u,v)-coverage, we 
find Vega is rotating less rapidly and with a smaller gravity darkening coefficient than previous interferometric results.  Our models are compatible with low photospheric macroturbulence and also consistent with the possible rotational period of $\sim$0.71~days recently reported based on magnetic field observations.  Our updated evolutionary analysis explicitly incorporates rapid rotation, finding Vega to have a mass of $2.15^{+0.10}_{-0.15}$\msun and an age $700^{-75}_{+150}$ Myrs, substantially older than previous estimates with errors dominated by lingering metallicity uncertainties ($Z=0.006^{+0.003}_{-0.002}$). 
\end{abstract}





\section{Introduction}
The nearby hot star Vega (spectral type A0) has been used as a photometric standard for millennia.  While Vega's relatively narrow spectral lines (\vsini$\sim$22 km/s) suggest slow rotation,  interferometric observations instead have established Vega to be a rapid rotator viewed near pole-on \citep{peterson2006,aufdenberg2006}, confirming suspicions of earlier spectroscopists \citep{gray1985,gray1988,gulliver1994}.  Rapid rotation should keep the surface material and stellar envelope well-mixed, leading to the conclusion that the observed sub-solar photospheric abundance represents the bulk composition.  Lower metallicity led to a revised lower mass estimate of $\sim$2.15\msun~for Vega and  increased  age $\sim$500~Myrs \citep{yoon2008,yoon2010}.

Based purely on spectroscopic analysis, Takeda and collaborators \citep{takeda2008a,takeda2008b} agree that Vega is rapidly rotating but with a preferred set of parameters at odds with the first-generation of interferometry results.  The parameters most discrepant are the rotational period and inclination angle, key values for modeling line profiles and understanding its evolutionary state.  Yoon et al. (2008, 2010) made the case that non-standard macroturbulence broadening of $\sim$10~km/s could accommodate both the observed line profile shapes and the interferometry results (Takeda\etal adopted 2~km/s microturbulence with no additional broadening).   \citet{hill2010} carried out a similar analysis and found intermediate results for best-fitting macroturbulence and inclination angle.  We refer here to this tension between models as Vega's {\em inclination controversy}, although one might alternatively refer to it as a {\em macroturbulence controversy}. 
  
A third observing method has recently shed new light on this touchstone system.  From analysis of circularly-polarized light, \citet{lignieres2009} found evidence for a weak magnetic field in Vega.  \citet{petit2010}
carried out Zeeman Doppler imaging, finding a detectable weak polar field concentration ($\sim$0.6 Gauss).  
Although the periodic signal is indeed weak, there is growing confidence after many years of observations that a persistent signal at 0.71$\pm$0.03 days \citep{petit2010,alina2012} represents the rotational period of Vega.  This period is substantially longer than expected from interferometry-based models (P$\sim0.5$--$0.6$~days) but compatible with the period range predicted using line profiles alone (P$\sim0.7$--$0.9$~days; Takeda\etal with no excess macroturbulence).

In this paper, we present extensive new interferometer observations of Vega using the Michigan Infrared Combiner (MIRC) on the Center for High Angular Resolution Astronomy (CHARA)  Array.  Our data have higher angular resolution than previous work and substantially improved Fourier coverage, allowing a robust estimate of internal and systematic errors.   Building on our recent imaging and modeling of other rapid rotators \citep{monnierscience2007, zhao2009, che2011}, we explore a wider range of gravity darkening prescriptions.  In short, we bring a new independent and critical look at the constraints interferometers can bring to Vega, particularly cognizant of parameter degeneracies and calibration systematics.

\section{Observations and Data Reduction}
\label{observations}
We have used CHARA  array in conjunction with the MIRC combiner to measure visibilities (\vis) and closure phases (CP) of Vega across the near-infrared H-band.  The CHARA Array was built and is operated by Georgia State University on Mt. Wilson, California.  CHARA is the longest baseline optical/infrared interferometer in the world with six fixed 1-m telescopes and a maximum baseline of 330m
\citep{theo2005}.

The MIRC image-plane combiner was used for all observations presented here.  Before 2011, MIRC was used to combine four telescope beams, allowing 6 \viss and 4 CP measurements at a time.  Following a major upgrade in 2011, MIRC now combines all six CHARA telescopes, resulting in up to 15 \viss and 20 CP measurements simultaneously.  MIRC splits the H band light ($\lambda_0=1.65\mu$m) into eight spectral channels ($\frac{\lambda}{\Delta\lambda}\sim$42), with absolute wavelength precision of  $\pm$0.25\% based on measures of $\iota$~Peg using the orbit of \citet{konacki2010}.
Further instrument details  can be found in a series of SPIE papers \citep{monnier_spie2004,monnier_spie2006,monnier_spie2010,che_spie2010,che_spie2012}.  
 
Using Fourier transform techniques, the \viss are measured, averaged and corrected for biases. The bispectrum is formed using the phases and amplitudes of three baselines that form a closed triangle \citep{monnier_phases2007}.   Amplitude calibration was performed using realtime flux estimates derived from choppers (before 2010) or through use of a beamsplitter following spatial filtering for improved performance \citep[after 2010;][]{che_spie2010}. Lastly, observations of reference calibrators throughout the night allowed for correction of time-variable 
 factors such as atmospheric coherence time, vibrations, differential dispersion, and birefringence in the beam train.   Additional pipeline details can be found in earlier papers \citep[e.g.,][]{monnierscience2007, zhao2009, che2011}.

For this work we have evolved our calibration model to better account for systematics.  Firstly, we include two types of calibration error for \viss -- multiplicative errors associated with the transfer function and additive errors associated with correcting biases at low fringe or bispectrum amplitude.  Based on calibrator studies, the former has been estimated to be 20\% (6.6\%) for 2007 (2012) data while the additive systematic error is $\Delta$\viss=$ 2\times10^{-4}$ (for both epochs).  For triple amplitudes (``T3amp''), the corresponding  multiplicative errors are 30\% (10\%) for 2007 (2012) and additive errors are $1\times10^{-5}$.  A detailed study by \citet{zhao2011} suggests CPs have an error floor of 1\arcdegg~for the observing modes adopted here.  To avoid our model fits being trapped by systematics, we also include two new types of CP errors associated with low signal-to-noise ratio (SNR) data near visibility null crossings.  Because correlated camera readout noise dominates the CP measurements at low SNR, we enforce minimum CP errors when the SNR$_{\rm T3amp}\simle1$.  In addition, we account for  finite time-averaging and spectral bandpass effects by including an error term proportional to $\Delta{\rm CP}_\lambda$ across each spectral channel.  Formally, these two terms are only important right at the null crossings and appear in the following noise floor formula:  $\sigma_{\rm CP} > {\rm MAX} ( \frac{30\arcdegg}{({\rm SNR}_{\rm T3amp})^{2}} , 0.2~ \Delta{\rm CP}_\lambda) $.  

Here we present data for Vega from 3 nights in 2007 (MIRC4) and 2 nights in 2012 (MIRC6).    Data on four additional nights of MIRC4 were recorded in 2007 and 2010  but were discarded due to calibration problems.   Table~\ref{table_obslog} includes detailed observing information including the calibrators and their adopted sizes; reduced data are available in OI-FITS format \citep{pauls2005} upon request.

Inspection of CPs show nearly all values to be at zero or 180\arcdegg~as expected for a point-symmetric intensity distribution.
Figure~1 shows the visibility data and the (u,v)-coverage (inset) for our datasets, split into three chunks of similar quantity: 2007, 2012 Jun 9 and 2012 Jun 13.   The data was azimuthally-averaged and compared to uniform disk and power-law limb-darkened disk
( $I = I_0 \mu^\alpha$ ) models.  As expected, the data are not consistent with a uniform disk and we find a best-fit limb-darkened diameter of 3.324~milliarcseconds (mas) with power law $\alpha\sim0.227$ -- more limb-darkened than expected for a non-rotating star ($\alpha\sim$0.11, Kurucz).  We note variation between epochs due to calibration errors.  Measuring precise limb-darkening requires controlling systematics at the few \% level, a goal for CHARA/MIRC but one that still proves challenging to attain during all observing conditions.  

Internal diagnostics demonstrate that the best calibration is from 2012 Jun 13. This night had the best flux calibration, the greatest on-source integration time, and also employed the maximum number of simultaneous telescopes.  The 2007 data, while extensive, were taken before our photometric channel upgrade and suffer from larger calibration errors.  For the detailed modeling in the next section, we limited our fits to the 2012 Jun 13 dataset.

\section{Modelling}
\label{analysis}
Our group has been a leader in the field of modeling rapid rotators, based largely on our unique and extensive interferometry data from CHARA/MIRC.  Our series of papers \citep{monnierscience2007, zhao2009, che2011} contains the first images of main-sequence stars beyond the Sun and has determined precise stellar parameters for rapidly rotating stars from early F to late B.  The basic physical model consists of a star with uniformly-rotating surface layers, distorted by centrifugal forces acting under point-gravity with a surface temperature following the gravity-darkening law $T = T_{\rm pole} (\frac{g}{g_{\rm pole}})^{\beta}$, where $g$ is the effective surface gravity. Based on our full dataset, \citet{che2011} argued that the observed gravity darkening deviates from the canonical value of $\beta=0.25$ advocated by
\citet{vonzeipel1924a,vonzeipel1924b} and instead we find empirically a lower characteristic value of $\beta=0.19$.   For our work here, we will again consider a range of possible $\beta$ coefficients.  Details of our calculations can be found in these earlier papers which followed the method of \citet{aufdenberg2006};  the full list of independent model parameters is shown at the top of Table~\ref{vega_tab}.

In order to carry out the full calculation, we had to assume a few physical parameters. We used a \citet{kurucz1979} grid (see http://kurucz.harvard.edu/grids.html) for a [Fe/H]=-0.5 (sub solar) plane-parallel atmosphere \citep[recent metallicity determinations by][]{yoon2010}. We also used the Hipparcos distance of 7.68~pc \citep{hipparcos2} and fixed the model mass to 2.15\msun~as recommended from \citet[][and consistent with final results presented here]{yoon2010}.

$\chi^2$-minimization was used to constrain our model parameters by employing the ensemble Markov Chain method described by \citet{emcee2012}, based on the affine-invariant strategy outlined in \citet{goodman2010}.  
We used 1000 walkers seeded by uncorrelated random distributions of the entire relevant parameter space.  The distribution reached convergence typically within 25 steps although we calculated 75 steps before freezing the distribution for error analysis.  The statistical weight of the \vis, T3amp and CP fits were reduced by the number of spectral channels ($=8$) because of strong internal correlations.  In addition, we down-weighted the contribution from the \viss and T3amp by an additional factor of two, since these quantities are not independent of each other (i.e., T3amps can be derived from the \vis).   The observed V-magnitude \citep[$0.03$~mag,][]{mermilliod1997} and H-magnitude \citep[$0.00$~mag,][]{kidger2003} photometry (adopting 5\% photometric errors to account for both measurement and zero-point uncertainties) was incorporated as a statistical {\em prior} during the Markov Chain calculation. Our final error bars combined the errors from the Markov Chain in quadrature with errors due to calibrator diameter uncertainty -- this was done by repeating the above Markov Chain for different assumed calibrator diameters drawn from the expected range of values.  These two error sources were often similar in magnitude, although some parameters have uncertainties dominated by one or the other.

Table~\ref{vega_tab} shows the full results of three separate parameter studies, labeled Model 1, 2, \& 3.  Because of the severe degeneracy between $\omega$ and $\beta$ for pole-on rapid rotators \citep[see discussion in ][]{zhao2009}, we chose to fix the gravity darkening coefficient for Model~1 to be $\beta=0.25$, consistent with the classical von Zeipel value allowing for comparison with previous work.  For Model 2, we fixed $\beta=0.19$ equal to the recommended value of \citet{che2011}.      Before describing Model~3, we first discuss the results of Models 1 \& 2.

Most of the model parameters are consistent (within uncertainties) between Models 1 \& 2. Model 1 prefers a slower rotational rate (longer periods) than Model 2, which was expected since a faster rotation state is needed to compensate for the weaker gravity darkening to maintain the center-to-limb intensity profile.  Note that the $\chi^2$ for Models 1 and 2 are practically identical ($\chi^2_\nu=0.89$ and $0.88$, respectively), showing the near perfect degeneracy in $\beta$--$\omega$ space for a pole-on system.    While our models do prefer low inclination angles ($i\sim4.5$\arcdegg), in agreement with earlier interferometric models \citep{peterson2006,aufdenberg2006,yoon2010}, the determination of our uncertainty is nearly 5$\times$ larger than those derived by earlier workers.  This is true for other parameters too -- our analysis takes a much more modest view  of parameter uncertainties which largely eliminates the strong conflict with spectroscopic analyses.  For instance, we find \vsini$=15\pm 4$\kms~(no macroturbulence) which is more compatible with the observed  22$\pm$2\kms.  The most notable statistically-significant disagreement with previous determinations is \wratio : for $\beta=$0.25, we find \wratio $=0.77\pm0.05$, not compatible with 0.93$\pm$0.02 \citep{peterson2006}, $0.91\pm0.03$ \citep{aufdenberg2006}, or 0.876$\pm$0.006 \citep{yoon2010}.  We speculate that the more limited datasets of these workers led to underestimates of systematic errors although we can not rule out certain physical explanations (time variability, spots, non-standard gravity darkening).

Constrained by only interferometry data, our parameters of Models 1 \& 2 span a larger range than earlier estimates, limiting our constraints on key stellar properties.  We can reduce our errors by including  constraints on the period \citep[0.71$\pm$0.03 days;][]{petit2010,alina2012} and \vsini (22$\pm$2\kms, Takeda et al. 2008ab) as statistical {\em priors} during the Markov Chain calculation.  We call Model 3 our ``Concordance Model," a set of parameters that agrees with CHARA/MIRC interferometry, the SED, \vsinii, and rotational period estimates.
Table~2 contains these results, showing much smaller error bars with only a minimal increase in normalized  $\chi^2$ (0.89 to 0.90).  In this model we allowed gravity darkening to be free and find $\beta=0.231\pm0.028$, a bit higher than the 0.19 preferred by \citet{che2011} but interestingly consistent with the re-parametrization of $\beta$ by \citet{espinosa2011}, who argue that $\beta$ depends on the rotation rate, matching 0.25 only for slow rotators  becoming smaller as the star spins up. 

As a side note, we found that the basic parameters of the Concordance Model can be determined simply and robustly without a complicated calculation. First, the projected equatorial diameter can be deduced from basic visibility fitting (see Figure~1) to be 3.32~mas.  This diameter can be turned into an equatorial velocity ($v_{\rm eq}=195$\kms) using the estimated period and distance. And finally, the inclination angle must be $\sim$6.5\arcdegg~to match the observed \vsinii.  Indeed, these ``back-of-the-envelope" estimates match quite closely the Concordance Model results in Table~2.

One way to view these results is to compare our parameters with the family of SED solutions outlined by \citet{takeda2008a}. Figure~2 shows our interferometric-based models (along with those of Peterson\etal and Aufdenberg\etal) plotted in $\omega$--inclination space next to Takeda's Models \#1-9\footnote{Takeda's curve was slightly shifted here to account for fact that we use 2.15\msun~instead of 2.3\msun:  $\sin i_{\rm new} =  \sqrt{\frac{M_{\rm Takeda}}{M_{\rm new}}}  \sin i_{\rm Takeda}$} .   \citet{takeda2008a} went further and used the line profile shapes to select best Models \#3, 4, 5, corresponding to $\sim$7\arcdegg ~inclination.  We see our Models 1 \& 2 are just consistent with the Takeda results at the $\sim$1-sigma level while our Concordance Model strongly selects Takeda Model \#5 as the optimal choice.  Indeed, this diagram further reinforces that a true concordance does exist between the SED and line profile fitting of Takeda, the putative rotational period from Petit\etal and the CHARA/MIRC interferometric observations in the near-infrared.

We end this section with a note of caution. The deviation from centro-symmetry on the surface of Vega is quite subtle, amounting to a pole offset of  just $\sim0.2$~milli-arcsecond, roughly 5$\times$ smaller than the fringe spacing from our longest baseline. Model-fitting different MIRC epochs can yield pole PAs as different as 90\arcdegg, and results from our best epoch (see Table~2) are discrepant with results from \citet{peterson2006}.   We have searched extensively for the explanation for the fragile constraints on the pole PA, including physical causes \citep[faint close-by companion, magnetic spots, non-radial pulsations; see also][]{rogers2012} and calibration-related problems (fringe cross talk, detector noise, bispectrum bias, time-averaging).   After an exhaustive series of tests, none of these hypotheses could convincingly explain the variations. We urge follow-up observations, especially at visible wavelengths where gravity darkening effects are strongest.  Fortunately, the conclusions from our work here depend mostly on \viss and not CPs, and the pole PA is not of paramount physical importance.

\section{Discussion and Conclusions}
\label{hrdiagram}
We can now use our modeling results to assess the evolutionary state of Vega.  We have used the most recent evolutionary tracks from the Geneva group that explicitly incorporates the effects of rapid rotation \citep[Georgy\etal 2013, submitted;][]{ekstrom2012}.  We considered metallicity range $Z=0.006^{+0.003}_{-0.002}$ appropriate for 
$[Fe/H]=-0.5$ under the range of currently considered chemical abundances of the Sun \citep{anders1989,asplund2005,asplund2009} -- note \citet{yoon2010} recommend $Z=0.009$ corresponding to the upper range we considered.  Because rapid rotation makes a star's position on the traditional H-R diagram ($L$ vs. $T_{\rm eff}$) to be viewing angle dependent \citep[see e.g.,][]{zhao2009}, we instead present stellar evolutionary tracks in units of total bolometric luminosity and stellar polar radius.  Figure~3 shows our modified Hertzsprung-Russell diagram for $Z=0.006$ including the effect of rotation. Our best models have \wratio$\sim$0.8 and we show these isochrones in the figure.  We conclude that Vega has a mass of $2.15^{+0.10}_{-0.15}$\msun~and age of $700^{-75}_{+150}$~million years for $Z=0.006^{+0.003}_{-0.002}$, with errors dominated by the metallicity assumption not random errors.  While our mass estimate is similar to those of \citet{yoon2010}, our age estimate is significantly higher due mostly to including the effect of rotation and less so because of the lower mean metallicity Z we have adopted.  

In conclusion, we have presented modeling of the photometric standard star Vega using new interferometric data from CHARA/MIRC. The large quantity and high angular resolution of our data allow for precise constraints on the geometry and surface temperatures of Vega. We find Vega rotating more slowly than previous interferometer results, consistent with the putative rotation period observed by \citet{alina2012} and compatible with the observed line profiles without excess macroturbulence. The differences with previous interferometry results could be from under-estimates of errors in earlier work or may suggest subtle deficiencies in the physical models. Our ``Concordance Model" and its placement on a new H-R diagram represent the best global model for Vega to date but there is still room for improvement.  In addition to confirmation of the rotation period through photometry, we recommend additional visible-light interferometry data spanning the first 3 visibility lobes with $<$5\% precision on V$^2$ to definitively establish the tilt angle of the pole and to pinpoint the true level of gravity darkening.

\acknowledgments {JDM thanks  Deane Peterson, Yoichi Takeda, and Pascal Petit for sharing their insights into Vega. The CHARA Array is currently funded by the National
Science Foundation through AST-1211929 and by the Georgia State University. Funding for the MIRC
combiner came from the University of Michigan and observations were
supported through NSF grants AST-0352723, AST-0707927,  and AST-1108963. STR acknowledges partial support from NASA grant NNH09AK731. This research has made use of the SIMBAD database, operated at CDS, Strasbourg, France, and NASA's Astrophysics Data System (ADS) Bibliographic Services.

{\it Facility:} \facility{CHARA (MIRC)}
}

\bibliographystyle{apj}


\begin{thebibliography}{41}
\expandafter\ifx\csname natexlab\endcsname\relax\def\natexlab#1{#1}\fi

\bibitem[Alina et al.(2012)]{alina2012} Alina, D., Petit, P., 
Ligni{\`e}res, F., et al.\ 2012, American Institute of Physics Conference 
Series, 1429, 82 

\bibitem[{{Anders} \& {Grevesse}(1989)}]{anders1989}
{Anders}, E. \& {Grevesse}, N. 1989, \gca, 53, 197

\bibitem[{{Asplund} {et~al.}(2005){Asplund}, {Grevesse}, \&
  {Sauval}}]{asplund2005}
{Asplund}, M., {Grevesse}, N., \& {Sauval}, A.~J. 2005, in Astronomical Society
  of the Pacific Conference Series, Vol. 336, Cosmic Abundances as Records of
  Stellar Evolution and Nucleosynthesis, ed. T.~G. {Barnes}, III \& F.~N.
  {Bash}, 25

\bibitem[{{Asplund} {et~al.}(2009){Asplund}, {Grevesse}, {Sauval}, \&
  {Scott}}]{asplund2009}
{Asplund}, M., {Grevesse}, N., {Sauval}, A.~J., \& {Scott}, P. 2009, \araa, 47,
  481

\bibitem[{{Aufdenberg} {et~al.}(2006){Aufdenberg}, {M{\'e}rand}, {Coud{\'e} du
  Foresto}, {Absil}, {Di Folco}, {Kervella}, {Ridgway}, {Berger}, {ten
  Brummelaar}, {McAlister}, {Sturmann}, {Sturmann}, \&
  {Turner}}]{aufdenberg2006}
{Aufdenberg}, J.~P., {M{\'e}rand}, A., {Coud{\'e} du Foresto}, V., et al. 2006, \apj, 645, 664

\bibitem[{{Barnes} {et~al.}(1978){Barnes}, {Evans}, \& {Moffett}}]{bem1978}
{Barnes}, T.~G., {Evans}, D.~S., \& {Moffett}, T.~J. 1978, \mnras, 183, 285

\bibitem[{{Bonneau} {et~al.}(2006){Bonneau}, {Clausse}, {Delfosse}, {Mourard},
  {Cetre}, {Chelli}, {Cruzal{\`e}bes}, {Duvert}, \& {Zins}}]{bonneau2006}
{Bonneau}, D., {Clausse}, J.-M., {Delfosse}, X., et al. 2006, \aap,
  456, 789

\bibitem[{{Che} {et~al.}(2012){Che}, {Monnier}, {Kraus}, {Baron}, {Pedretti},
  {Thureau}, \& {Webster}}]{che_spie2012}
{Che}, X., {Monnier}, J.~D., {Kraus}, et al. 2012, in Society of Photo-Optical
  Instrumentation Engineers (SPIE) Conference Series, in press

\bibitem[{{Che} {et~al.}(2010){Che}, {Monnier}, \& {Webster}}]{che_spie2010}
{Che}, X., {Monnier}, J.~D., \& {Webster}, S. 2010, in Society of Photo-Optical
  Instrumentation Engineers (SPIE) Conference Series, Vol. 7734

\bibitem[{{Che} {et~al.}(2011){Che}, {Monnier}, {Zhao}, {Pedretti}, {Thureau},
  {M{\'e}rand}, {ten Brummelaar}, {McAlister}, {Ridgway}, {Turner}, {Sturmann},
  \& {Sturmann}}]{che2011}
{Che}, X., {Monnier}, J.~D., {Zhao}, M., et al. 2011, \apj, 732, 68

\bibitem[{{Ekstr{\"o}m} {et~al.}(2012){Ekstr{\"o}m}, {Georgy}, {Eggenberger},
  {Meynet}, {Mowlavi}, {Wyttenbach}, {Granada}, {Decressin}, {Hirschi},
  {Frischknecht}, {Charbonnel}, \& {Maeder}}]{ekstrom2012}
{Ekstr{\"o}m}, S., {Georgy}, C., {Eggenberger}, P., et al. 2012, \aap, 537, A146

\bibitem[{{Espinosa Lara} \& {Rieutord}(2011)}]{espinosa2011}
{Espinosa Lara}, F. \& {Rieutord}, M. 2011, \aap, 533, A43

\bibitem[{{Foreman-Mackey} {et~al.}(2012){Foreman-Mackey}, {Hogg}, {Lang}, \&
  {Goodman}}]{emcee2012}
{Foreman-Mackey}, D., {Hogg}, D.~W., {Lang}, D., \& {Goodman}, J. 2012, ArXiv
  e-prints

\bibitem[{{Goodman} \& {Weeare}(2010)}]{goodman2010}
{Goodman}, J. \& {Weeare}, J. 2010, Communications in Applied Mathematics and
  Computational Science, 5, 65

\bibitem[{{Gray}(1985)}]{gray1985}
{Gray}, R.~O. 1985, \jrasc, 79, 237

\bibitem[{{Gray}(1988)}]{gray1988}
---. 1988, \jrasc, 82, 336

\bibitem[{{Gulliver} {et~al.}(1994){Gulliver}, {Hill}, \&
  {Adelman}}]{gulliver1994}
{Gulliver}, A.~F., {Hill}, G., \& {Adelman}, S.~J. 1994, \apjl, 429, L81

\bibitem[Hill et al.(2010)]{hill2010} Hill, G., Gulliver, A.~F., 
\& Adelman, S.~J.\ 2010, \apj, 712, 250 

\bibitem[{{Kidger} \& {Mart{\'{\i}}n-Luis}(2003)}]{kidger2003}
{Kidger}, M.~R. \& {Mart{\'{\i}}n-Luis}, F. 2003, \aj, 125, 3311

\bibitem[{{Konacki} {et~al.}(2010){Konacki}, {Muterspaugh}, {Kulkarni}, \&
  {He{\l}miniak}}]{konacki2010}
{Konacki}, M., {Muterspaugh}, M.~W., {Kulkarni}, S.~R., \& {He{\l}miniak},
  K.~G. 2010, \apj, 719, 1293

\bibitem[{{Kurucz}(1979)}]{kurucz1979}
{Kurucz}, R.~L. 1979, \apjs, 40, 1

\bibitem[{{Ligni{\`e}res} {et~al.}(2009){Ligni{\`e}res}, {Petit}, {B{\"o}hm},
  \& {Auri{\`e}re}}]{lignieres2009}
{Ligni{\`e}res}, F., {Petit}, P., {B{\"o}hm}, T., \& {Auri{\`e}re}, M. 2009,
  \aap, 500, L41

\bibitem[{{Mermilliod} {et~al.}(1997){Mermilliod}, {Mermilliod}, \&
  {Hauck}}]{mermilliod1997}
{Mermilliod}, J.-C., {Mermilliod}, M., \& {Hauck}, B. 1997, \aaps, 124, 349

\bibitem[{{Monnier}(2007)}]{monnier_phases2007}
{Monnier}, J.~D. 2007, \nar, 51, 604

\bibitem[{{Monnier} {et~al.}(2010){Monnier}, {Anderson}, {Baron}, {Berger},
  {Che}, {Eckhause}, {Kraus}, {Pedretti}, {Thureau}, {Millan-Gabet}, {Ten
  Brummelaar}, {Irwin}, \& {Zhao}}]{monnier_spie2010}
{Monnier}, J.~D., {Anderson}, M., {Baron}, F., et al. 2010, in Society of
  Photo-Optical Instrumentation Engineers (SPIE) Conference Series, Vol. 7734

\bibitem[{{Monnier} {et~al.}(2004){Monnier}, {Berger}, {Millan-Gabet}, \& {ten
  Brummelaar}}]{monnier_spie2004}
{Monnier}, J.~D., {Berger}, J., {Millan-Gabet}, R., \& {ten Brummelaar}, T.~A.
  2004, in the Society of Photo-Optical Instrumentation Engineers
  (SPIE) Conference, Vol. 5491, 1370

\bibitem[{{Monnier} {et~al.}(2006){Monnier}, {Pedretti}, {Thureau}, {Berger},
  {Millan-Gabet}, {ten Brummelaar}, {McAlister}, {Sturmann}, {Sturmann},
  {Muirhead}, {Tannirkulam}, {Webster}, \& {Zhao}}]{monnier_spie2006}
{Monnier}, J.~D., {Pedretti}, E., {Thureau}, et al. 2006, in
 the Society of Photo-Optical Instrumentation Engineers (SPIE)
  Conference, Vol. 6268

\bibitem[{{Monnier} {et~al.}(2007){Monnier}, {Zhao}, {Pedretti}, {Thureau},
  {Ireland}, {Muirhead}, {Berger}, {Millan-Gabet}, {Van Belle}, {ten
  Brummelaar}, {McAlister}, {Ridgway}, {Turner}, {Sturmann}, {Sturmann}, \&
  {Berger}}]{monnierscience2007}
{Monnier}, J.~D., {Zhao}, M., {Pedretti}, E., et al. 2007, Science, 317, 342

\bibitem[{{Pauls} {et~al.}(2005){Pauls}, {Young}, {Cotton}, \&
  {Monnier}}]{pauls2005}
{Pauls}, T.~A., {Young}, J.~S., {Cotton}, W.~D., \& {Monnier}, J.~D. 2005,
  \pasp, 117, 1255

\bibitem[{{Peterson} {et~al.}(2006){Peterson}, {Hummel}, {Pauls}, {Armstrong},
  {Benson}, {Gilbreath}, {Hindsley}, {Hutter}, {Johnston}, {Mozurkewich}, \&
  {Schmitt}}]{peterson2006}
{Peterson}, D.~M., {Hummel}, C.~A., {Pauls}, T.~A., et al. 2006, \nat, 440,
  896

\bibitem[{{Petit} {et~al.}(2010){Petit}, {Ligni{\`e}res}, {Wade},
  {Auri{\`e}re}, {B{\"o}hm}, {Bagnulo}, {Dintrans}, {Fumel}, {Grunhut},
  {Lanoux}, {Morgenthaler}, \& {van Grootel}}]{petit2010}
{Petit}, P., {Ligni{\`e}res}, F., {Wade}, G.~A., et al. 2010, \aap, 523, A41

\bibitem[{{Rogers} {et~al.}(2012){Rogers}, {Lin}, \& {Lau}}]{rogers2012}
{Rogers}, T.~M., {Lin}, D.~N.~C., \& {Lau}, H.~H.~B. 2012, \apjl, 758, L6





\bibitem[{{Takeda} {et~al.}(2008a){Takeda}, {Kawanomoto}, \&
  {Ohishi}}]{takeda2008a}
Takeda, Y., Kawanomoto, 
Y.,  \& Ohishi, N.\ 2008a, \apj, 678, 446


\bibitem[Takeda et al.(2008b)]{takeda2008b} 
---.\ 2008b, Contributions of the Astronomical Observatory Skalnate Pleso, 38, 157 


\bibitem[{{ten Brummelaar} {et~al.}(2005){ten Brummelaar}, {McAlister},
  {Ridgway}, {Bagnuolo}, {Turner}, {Sturmann}, {Sturmann}, {Berger}, {Ogden},
  {Cadman}, {Hartkopf}, {Hopper}, \& {Shure}}]{theo2005}
{ten Brummelaar}, T.~A., {McAlister}, H.~A., {Ridgway}, S.~T., et al. 2005, \apj, 628, 453

\bibitem[{{van Leeuwen}(2007)}]{hipparcos2}
{van Leeuwen}, F. 2007, \aap, 474, 653

\bibitem[{{von Zeipel}(1924{\natexlab{a}})}]{vonzeipel1924a}
{von Zeipel}, H. 1924{\natexlab{a}}, \mnras, 84, 665

\bibitem[{{von Zeipel}(1924{\natexlab{b}})}]{vonzeipel1924b}
---. 1924{\natexlab{b}}, \mnras, 84, 684

\bibitem[{{Yoon} {et~al.}(2010){Yoon}, {Peterson}, {Kurucz}, \&
  {Zagarello}}]{yoon2010}
{Yoon}, J., {Peterson}, D.~M., {Kurucz}, R.~L., \& {Zagarello}, R.~J. 2010,
  \apj, 708, 71

\bibitem[{{Yoon} {et~al.}(2008){Yoon}, {Peterson}, {Zagarello}, {Armstrong}, \&
  {Pauls}}]{yoon2008}
{Yoon}, J., {Peterson}, D.~M., {Zagarello}, R.~J., {Armstrong}, J.~T., \&
  {Pauls}, T. 2008, \apj, 681, 570

\bibitem[{{Zhao} {et~al.}(2011){Zhao}, {Monnier}, {Che}, {Pedretti}, {Thureau},
  {Schaefer}, {Ten Brummelaar}, {M{\'e}rand}, {Ridgway}, {McAlister}, {Turner},
  {Sturmann}, {Sturmann}, {Goldfinger}, \& {Farrington}}]{zhao2011}
{Zhao}, M., {Monnier}, J.~D., {Che}, et al. 2011, \pasp, 123, 964

\bibitem[{{Zhao} {et~al.}(2009){Zhao}, {Monnier}, {Pedretti}, {Thureau},
  {M{\'e}rand}, {ten Brummelaar}, {McAlister}, {Ridgway}, {Turner}, {Sturmann},
  {Sturmann}, {Goldfinger}, \& {Farrington}}]{zhao2009}
{Zhao}, M., {Monnier}, J.~D., {Pedretti}, et al.
  2009, \apj, 701, 209

\end{thebibliography}

\clearpage

\begin{deluxetable}{lcccl}
\tabletypesize{\scriptsize}
\tablecaption{CHARA/MIRC log for Vega observations
\label{table_obslog}}
\tablewidth{0pt}
\tablehead{  \colhead{Date} &  \colhead{Interferometer}  &
\colhead{Number of} & \colhead{Number of} & \colhead{Calibrator} \\ 
\colhead{(UT)} & \colhead{(Configuration)} & 
\colhead{\viss} & \colhead{Closure Phases} & \colhead{Information} }
\startdata
2007 Jul 5& S1--E1--W1--W2 & 168 & 104 & $\sigma$~Cyg\tablenotemark{a}, $\Upsilon$~Peg\tablenotemark{b}\\
2007 Jul 8 & S1--E1--W1--W2 & 96 & 64 & $\gamma$~Lyr\tablenotemark{c} $\Upsilon$~Peg  \\
2007 Jul 13 & S1--E1--W1--W2 & 144 & 96 & $\sigma$~Cyg  \\
2012 Jun 9 & S2--S1--E1--E2--W2 & 200 & 144 & HD~167304\tablenotemark{d} \\
2012 Jun 13 & W1--S2--S1--E1--E2--W2 & 560 & 640 &$\gamma$~Lyr \\
\enddata   
\tablenotetext{a}{Adopted $\sigma$~Cyg UD diameter $0.54\pm0.02$~mas \citep{bem1978}}
\tablenotetext{b}{Adopted $\Upsilon$~Peg UD diameter $0.99\pm0.02$~mas (new CHARA/MIRC measurement)}
\tablenotetext{c}{Adopted $\gamma$~Lyr UD diameter $0.737\pm0.015$~mas based on independent measurements by CHARA/MIRC (UD$_H=0.723\pm0.025$~mas) and CHARA/PAVO  (UD$_H=0.744\pm0.019$~mas derived from UD$_{\rm LDD}=0.755\pm0.019$~mas)}
\tablenotetext{d}{Adopted HD~167304 UD diameter $0.69\pm0.05$~mas \citep{bonneau2006}}
\end{deluxetable}

\begin{deluxetable}{lccc}
\tabletypesize{\scriptsize}
\tablecaption{Modeling Results for Vega}
\tablewidth{0pt}
\tablehead{
\colhead{Model Parameters\tablenotemark{a}}
&\colhead{Model 1}
&\colhead{Model 2}
&\colhead{Model 3}\\
&\colhead{$\beta=0.25$}
&\colhead{$\beta=0.19$}
&\colhead{Concordance\tablenotemark{b}} 
}
\startdata
Inclination (\arcdegg) & $   4.5\pm   1.3$ & $   4.0\pm   1.5$ & $   6.2\pm   0.4$\\
Pole position angle (East of North \arcdegg) & $-57\pm7$ & $-57\pm6$ & $-58\pm6$\\
\tpol (\K) & $10120\pm140$ & $10130\pm140$ & $10070\pm90$\\
\rpol (\rsun) & $  2.42\pm  0.05$ & $  2.31\pm  0.06$ & $ 2.418\pm 0.012$\\
\rpol (mas) & $  1.47\pm  0.03$ & $  1.40\pm  0.04$ & $ 1.465\pm 0.007$\\
\wratio & $  0.77\pm  0.05$ & $  0.86\pm  0.04$ & $ 0.774\pm 0.012$\\
$\beta$ & $  0.25$ (FIXED) & $  0.19$ (FIXED) & $ 0.231\pm 0.028$\\
\hline
\hline
\colhead{Derived Parameters} & & & \\
\hline
\teq (\K) & $8870\pm200$ & $8740\pm190$ & $8910\pm130$\\
\req (\rsun) & $ 2.726\pm 0.007$ & $ 2.728\pm 0.007$ & $ 2.726\pm 0.006$\\
\req (mas) & $ 1.651\pm 0.004$ & $ 1.652\pm 0.004$ & $ 1.651\pm 0.004$\\
v / v$_{\rm crit}$ & $  0.58\pm  0.05$ & $  0.68\pm  0.05$ & $ 0.581\pm 0.012$\\
Bolometric luminosity \lbol (\lsun)  & $  47.1\pm   2.7$ & $  44.8\pm   2.6$ & $  47.2\pm   2.0$\\
Apparent luminosity \lapp(\lsun)    & $  58.8\pm   2.7$ & $  58.8\pm   2.7$ & $  58.4\pm   2.2$\\
Apparent effective temperature \tapp(\K)   & $9680\pm110$ & $9670\pm110$ & $9660\pm90$\\
Surface-averaged \teff(\K)   & $9350\pm110$ & $9310\pm110$ & $9360\pm90$\\
\vsini (\kms )      & $  15.1\pm   3.6$ & $  16.2\pm   4.8$ & $  21.3\pm   1.2$\tablenotemark{b}\\
Rotational period (days)      & $  0.71\pm  0.07$ & $  0.59\pm  0.06$ & $  0.71\pm  0.02$\tablenotemark{b}\\
Model $\emph{V}$ magnitude\tablenotemark{a}         & $ 0.032\pm 0.028$ & $ 0.032\pm 0.028$ & $ 0.035\pm 0.023$\\
Model $\emph{H}$ magnitude\tablenotemark{a}         & $ 0.029\pm 0.013$ & $ 0.030\pm 0.013$ & $ 0.031\pm 0.011$\\
Model mass (\msun)\tablenotemark{c}        &\multicolumn{3}{c}{Joint Result: $2.15^{+0.10}_{-0.15}$}\\
Model age (Myrs)\tablenotemark{c}    &\multicolumn{3}{c}{Joint Result: $700^{-75}_{+150}$}\\
\hline
\hline
\colhead{Summary of $\chi^{2}$ Results}      &               &                                                       &                                               \\
\hline
Total $\chi^2_{\nu}$ ($N_{\rm DATA}=1842$)     & $  0.89$ & $  0.88$ & $  0.90$ \\
Vis$^2$ $\chi^2_{\nu}$ ($N_{\rm DATA}=560$)  & $  1.45$ & $  1.45$ & $  1.44$ \\
T3amp $\chi^2_{\nu}$ ($N_{\rm DATA}=640$) & $  0.57$ & $  0.57$ & $  0.58$ \\
CP $\chi^2_{\nu}$ ($N_{\rm DATA}=640$)  & $  0.83$ & $  0.81$ & $  0.84$ \\
\hline
\enddata
\tablenotetext{a}{Other parameters: distance 7.68~pc \citep{hipparcos2}, V mag 0.03$\pm$0.05 \citep{mermilliod1997}, H mag 0.00$\pm$0.05 \citep{kidger2003}, $[Fe/H]=-0.5$ \citep{yoon2008}.}
\tablenotetext{b}{The Concordance Model incorporated the observed \vsini$=22\pm$2\kms \citep{takeda2008a}  and period estimate 0.71$\pm$0.03\,days \citep{petit2010,alina2012} into the fit as {\em priors}.}
\tablenotetext{c}{Geneva stellar evolutionary tracks \citep[Georgy\etal\,2013, submitted;][]{ekstrom2012} were used assuming \wratio$=0.8$ and covering the range $Z=0.006^{+0.003}_{-0.002}$.}
\label{vega_tab}
\end{deluxetable}

\clearpage

\begin{figure}[hbt]
\begin{center}
\includegraphics[angle=90,width=6in]{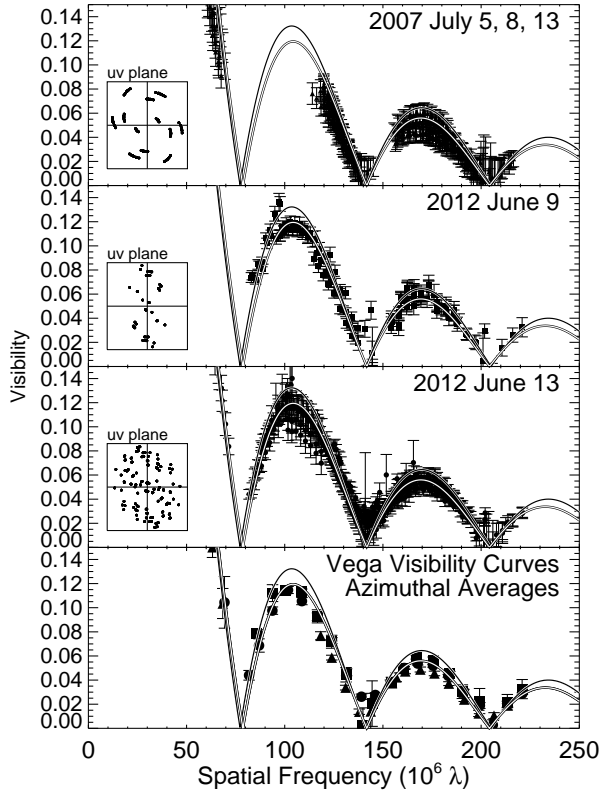}
\figcaption{\footnotesize  This figure shows three epochs of visibility observations using CHARA/MIRC.  The data are not consistent with a uniform disk fit (top line, 3.26~mas). The power-law limb-darkened disk fit (bottom line, 3.32~mas, $\alpha=0.23$) shows twice the level of expected limb-darkening, suggesting strong gravity darkening.  The (u,v) coverage for each epoch is shown in each panel as an inset box ($\pm$350~meters). 
\label{fig1}}
\end{center}
\end{figure}

\begin{figure}[hbt]
\begin{center}
\includegraphics[angle=90,width=6in]{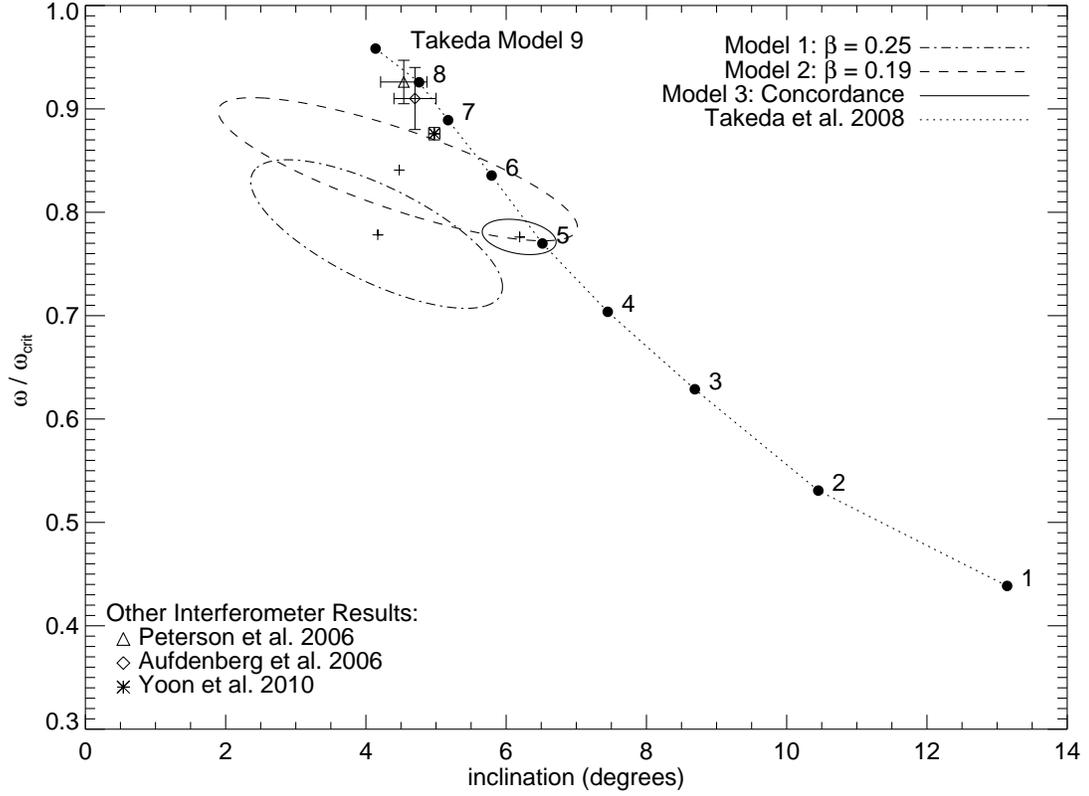}
\figcaption{\footnotesize 
This figure shows the \wratio~ vs. inclination angle results for our three models from Table~2. The ellipses show the 1-sigma confidence intervals. We also include the previous interferometric modelling results from \citet{peterson2006}, \citet{aufdenberg2006}, and \citet{yoon2010} as well as the family of solutions (Takeda Models \#1--9) presented in \citet[][adjusted here for $M_\ast=$2.15\msun]{takeda2008a}. 
 \label{fig2}}
\end{center}
\end{figure}

\begin{figure}[hbt]
\begin{center}
\includegraphics[angle=90,width=6in]{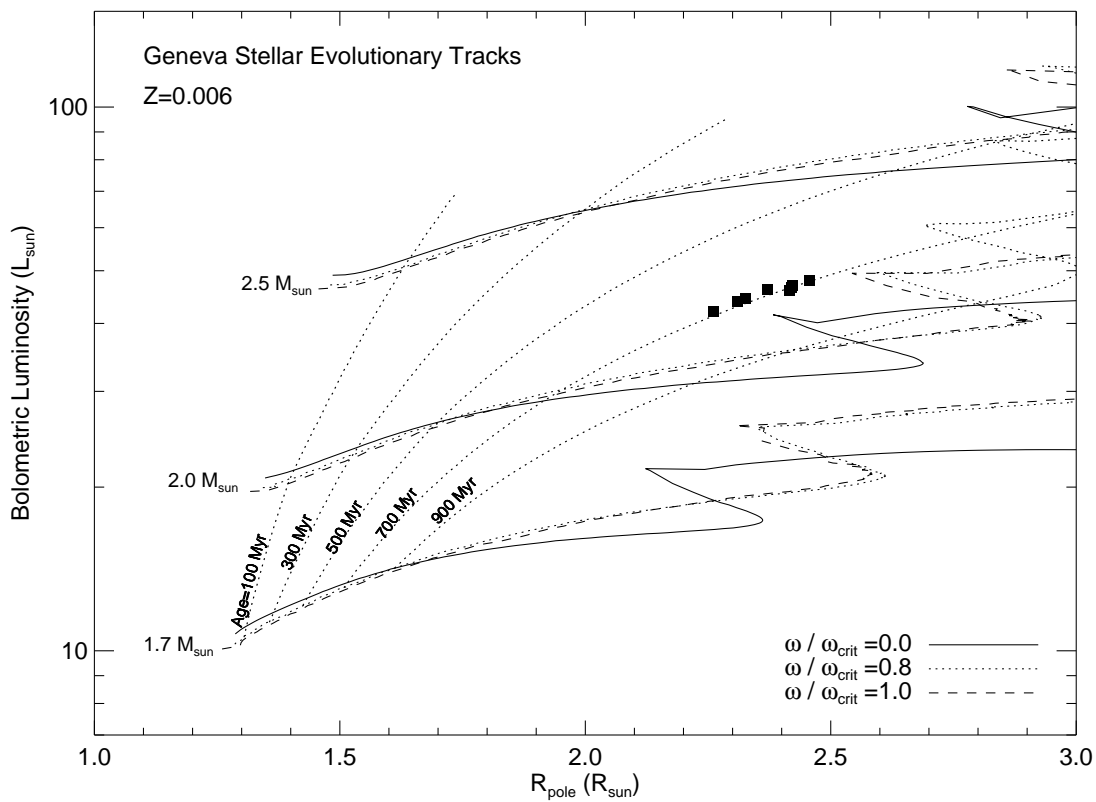}
\figcaption{\footnotesize  This modified Hertzsprung-Russell (H-R) diagram plots the total bolometric luminosity vs. the stellar polar radius for our best-fit Models 1, 2, \& 3 (squares), including the effect of the calibrator size uncertainty ($\pm1\sigma$).
The stellar evolutionary tracks are based on the most recent Geneva models that incorporate rotation \citep[Georgy\etal 2013, submitted;][]{ekstrom2012}, and we show isochrones for \wratio$ = 0.8$ (dotted line).
\label{fig3}}
\end{center}
\end{figure}

\end{document}